\documentclass[twocolumn,aps,superscriptaddress,showpacs]{revtex4}
\usepackage{amssymb}
\usepackage{amsmath}
\usepackage{graphicx}
\usepackage[normalem]{ulem}
\usepackage[dvips]{color}
\usepackage{epsfig}

\newcommand{\be}{\begin{eqnarray}}
\newcommand{\ee}{\end{eqnarray}}

\def\lsim{\mathrel{\rlap{\lower3pt\hbox{\hskip1pt$\sim$}}
     \raise1pt\hbox{$<$}}} 
\def\gsim{\mathrel{\rlap{\lower3pt\hbox{\hskip1pt$\sim$}}
     \raise1pt\hbox{$>$}}} 

\def\la{\langle}\def\ra{\rangle}

\def\bi{\bibitem}

\begin{document}
\title{Half-Skyrmions, Tensor Forces and Symmetry Energy \\
in Cold Dense Matter}
\author{Hyun Kyu Lee}
\affiliation{Department of Physics, Hanyang University, Seoul 133-791, Korea}
\author{Byung-Yoon Park}
\affiliation{Department of Physics, Chungnam National University, Daejon 305-764, Korea}
\author{Mannque Rho}
\affiliation{Institut de Physique Th\'eorique,  CEA Saclay, 91191 Gif-sur-Yvette C\'edex, France \&\\
Department of Physics, Hanyang University, Seoul 133-791, Korea}

\begin{abstract}
In a previous article, the 4D half-skyrmion (or 5D dyonic salt) structure of dense baryonic matter described in crystalline configuration in the large $N_c$ limit was shown to have nontrivial consequences on how anti-kaons behave in compressed nuclear matter with a possible implication on an ``ice-9" phenomenon of deeply bound kaonic matter and condensed kaons in compact stars. We extend the analysis to make a further prediction on the scaling properties of hadrons that have a surprising effect on the nuclear tensor forces, the symmetry energy and hence on the phase structure at high density.  We treat this problem relying on certain topological structure of chiral solitons. Combined with what can be deduced from hidden local symmetry for hadrons in dense medium and the ``soft" dilatonic degree of freedom associated with the trace anomaly of QCD, we uncover a novel structure of chiral symmetry in the ``supersoft" symmetry energy that can influence the structure of neutron stars. .
 \end{abstract}

\date{\today}

\newcommand\sect[1]{\emph{#1}---}

\maketitle
\section{Introduction}
When $A$ skyrmions with $A\rightarrow \infty$ are put on an FCC crystal lattice and squeezed to simulate dense baryonic matter, the skyrmion in the system is found to fractionize into two half-skyrmions at a density $n_{1/2}\sim xn_0$ with $x>1$ where $n_0$ is the normal nuclear matter density~\cite{PKR}. The matter made up of the half-skyrmions is characterized by the vanishing quark condensate $\la\bar{q}q\ra \propto {\rm Tr}U =0$ and the non-vanishing pion decay constant $f_\pi\neq 0$, whereas the lower-density skyrmion state has both $\la\bar{q}q\ra\neq 0$ and $f_\pi\neq 0$ symptomatic of chiral symmetry spontaneous breaking. A similar structure was found~\cite{RSZ} with instantons on an FCC crystal fractionizing into two half-instantons (or dyons) in 5D Yang-Mills theory in the gravity sector that arises in holographic QCD~\cite{SS}. What distinguishes the instanton baryon in the bulk gravity sector from the skyrmion baryon in the boundary gauge sector is that the former involves an infinite tower of vector mesons, so the physics of highly dense matter will be more efficiently accessed with higher-energy degrees of freedom incorporated.

It was found in \cite{PKR} that the mass of an anti-kaon propagating in dense matter undergoes a more propitious decrease as skyrmions fractionize into half-skrymions. It was suggested there that this behavior could trigger a deeply bound kaonic state in nuclear matter, a sort of ``ice-9" phenomenon~\cite{ice9} and kaon condensation in neutron star matter at a lower density than thought previously. In this paper, based on certain generic structure of the dense skrymion matter, we will make several further predictions that have not been made before, specifically on the structure of the tensor forces operative in dense matter at $n >n_0$, its ramifications on the symmetry energy of asymmetric nuclear matter and consequently on the structure of compact stars.

\section{The Model}
The model we shall use to study cold dense baryonic matter put in crystals is the two-flavor Skyrme model with the quartic Skyrme term, supplemented with a dilaton scalar $\chi$, analyzed in \cite{PKR},
\begin{eqnarray}
{\cal L}_{sk}
&=& \frac{f_\pi^2}{4} \left(\frac{\chi}{f_\chi}\right)^2
{\rm Tr} (L_\mu L^\mu) + \frac{1}{32e^2}{\rm Tr} [L_\mu, L_\nu]^2\nonumber\\
&& +\frac{f_\pi^2}{4}\left(\frac{\chi}{f_\chi}\right)^3
{\rm Tr}{\cal M} (U+U^\dagger-2),
\nonumber \\
&&+\frac{1}{2}\partial_\mu \chi\partial^\mu \chi + V(\chi)\label{skyrme-lag}
\end{eqnarray}
where $V(\chi)$ is the potential that encodes the trace anomaly involving the
soft dilaton (precisely defined in \cite{LR-dilaton}), $ L_{\mu} = U^{\dagger} \partial_{\mu} U $, with
$U$ the chiral field taking values in $SU(2)$ and $f_\chi$ is the vev of $\chi$.

In extracting the information needed for describing hadrons, both bosonic and baryonic, in dense medium, we will consider the Lagrangian (\ref{skyrme-lag}) -- without the scalar field -- as ``gauge equivalent" to the hidden local symmetry Lagrangian with the $U(2)$ multiplet $\rho$ and $\omega$~\cite{hls}. More precisely one can think of (\ref{skyrme-lag}) as resulting from integrating out all vector mesons from a Lagrangian that contains an infinite tower of hidden local fields such as, e.g., in the hQCD model of Sakai and Sugimoto~\cite{SS} based on string theory or deconstructed bottom-up from low-energy current algebra term~\cite{SS-son}. Thus although we simulate the skyrmion matter with (\ref{skyrme-lag}), we will be able to make statements on the vector mesons which are not explicit degrees of freedom in the Lagrangian. While (\ref{skyrme-lag}) has been widely studied both for the elementary nucleon and for many-nucleon systems in the literature, we find that it has a surprising feature that has so far remained unexposed, particularly in many-nucleon systems. It makes certain novel predictions based on topological structure of the soliton contained in the model, which has some dramatic -- hitherto unsuspected -- effects on compressed baryonic matter relevant to compact stars.

Over-simplified as it may appear, the Lagrangian (\ref{skyrme-lag}) could be justified as a candidate effective field theory for the physics of dense matter on several grounds. The first term without the coupling to the dialton field is of course the current algebra term rigorously valid at very low energy. The second term, called Skyrme term, often considered as {\it ad hoc}, is also justifiable. Although it is in principle not the only term that can appear in the chiral Lagrangian at fourth order in derivative, chiral perturbation calculations of $\pi$-$\pi$ scattering show that it is the dominant term, with other terms essentially canceling out~\cite{diakonov}. This contrasts with the linear sigma model in which three other four-derivatives terms combined together are found to destabilize the soliton and make it collapse to a point. Another information on the Skyrme term comes from holographic QCD (hQCD) constructed by Sakai and Sugimoto~\cite{SS} in which, present with the infinite tower of vector mesons, it turns out to be the only term quartic in derivatives acting on the pion field. In fact, the coefficient $1/e^2$ in the hQCD Lagrangian is precisely fixed by $\frac{1}{216\pi^2}\lambda N_c$ where $\lambda$ is the 't Hooft constant and $N_c$ is the number of colors. Surprisingly this coefficient comes out numerically very close to what has been found in the Skyrme model\cite{footnote0}.

More generally, as mentioned above, one may consider the Skyrme model as resulting from integrating out {\em all} vector degrees of freedom from the infinite tower of hidden local gauge fields that arises as emergent or ``deconstructed" bottom-up from the current algebra or reduced top-down from the 5D YM theory in hQCD~\cite{footnote-integrating}. The Skyrme term may be taken to encapsulate short-distance degrees of freedom that include quarks and gluons. Of course, there is no reason why one can simply stop at the quartic order in derivative. In general with higher order terms, the number of parameters increases rapidly although within the holographic model \`a la Sakai-Sugimoto, the situation is somewhat ameliorated.

In using (\ref{skyrme-lag}),  we choose to pick, as advocated in \cite{NRZ}, the parameters of the Lagrangian determined in the meson sector, not taken as free parameters as has been usually done in the literature. We thus take the pion decay constant to be given by $f_\pi\approx 93$ MeV, and the Skyrme term constant $1/e^2$ as given by the lowest mass scale integrated out, namely, the vector meson mass, $m_\rho\approx \sqrt{2} f_\pi e$. Of course, the nucleon mass comes out too high with these constants, say, $\sim 1500$ MeV. But this is the mass given at the leading order, ${\cal O}(N_c)$. As such, this high value should not worry us. In fact, the next order $({\cal O}(N_c^0))$ term, i.e., the Casimir term, -- which is difficult to calculate precisely -- is estimated to be $\sim -500$ MeV\cite{footnote1}.

In applying the Skyrme Lagrangian to many-nucleon systems, one glaring defect is the missing scalar degree of freedom that plays a key role in nuclear dynamics. This was recognized already in 1991 when the scaling relations were first written down~\cite{BR91}. Unlike the scalar $\sigma$ in the linear sigma model which does not support stable nuclear matter, the $\chi$ field in (\ref{skyrme-lag}) is a  chiral scalar locked to the chiral condensate $\la\bar{q}q\ra$. How to introduce the scalar field $\chi$ in chiral Lagrangians -- which is not at all trivial -- was discussed in \cite{LR-dilaton}. What is needed for our purpose is the ``soft dilaton" figuring in the trace anomaly of QCD whose condensate is locked to the chiral condensate.  This scalar mode can be thought of as representing the vibrational mode (i.e., the Casimir effect) -- which is subleading in $N_c$ -- missing in the Skyrme model mentioned above. This interpretation is supported by the result obtained in \cite{PRV-dilaton} where it is found that for large dilaton mass $m_\chi\gsim 1.3$ GeV, the soliton mass $M_{\rm sol}$ comes out to be $\gsim 1.4$ GeV whereas for $m_\chi < 1$ GeV, it is $\lsim 1$ GeV. Thus, the role of the soft dilaton is equivalent to that of the ${\cal O}(N_c^0)$ Casimir effect. Furthermore multiplying the Wess-Zumino term $\propto \omega\cdot B$ with $\chi^n$ with $n\gsim 2$ also simulates {\em by fiat} the property of ``vector manifestation" of hidden local symmetry that dictates that the vector meson coupling to the pions vanish $\propto \la\bar{q}q\ra$ as the chiral transition point is approached in the chiral limit~\cite{HY:PR}. This property could be more efficiently -- and should be -- addressed in the framework of hidden local symmetry Lagrangian with the hidden local fields present together with the pions and the dilaton. By putting the $\omega$ meson together with the $\rho$ meson in an $U(2)$ multiplet, it can counterbalance the possible over-binding by the scalar $\chi$ field as discussed below\cite{footnote1p}\cite{footnoteattempt}.
\section{Half-Skyrmion Crystal}
We now specify the features of the dense matter constructed with the Lagrangian (\ref{skyrme-lag}) that will be exploited in this paper.

The method we use to describe dense baryonic matter with the chiral Lagrangian (\ref{skyrme-lag}) -- valid at large $N_c$ -- is to put multi skyrmions on crystal lattice as pioneered by Klebanov~\cite{klebanov} and squeeze the system to simulate density. The most recent review on this approach is found in \cite{byp-vv}. Here we will rely on the results obtained in \cite{PKR} with the Lagrangian (\ref{skyrme-lag}) by putting skyrmions on an FCC crystalline. With the parameters $f_\pi$ and $1/e^2$ fixed as described above, there is only one parameter remaining to be fixed, namely, the mass of the dilaton $m_\chi$. At present, there is no clear information on $m_\chi$, both experimentally and theoretically. There is a great deal of controversy on scalar mesons involving both quarkonic and gluonic configurations. In the absence of a better  guidance, we will take two values that we consider reasonable for our problem. One is $\sim 600$ MeV which corresponds, roughly, to the lowest scalar with a broad width listed in PDG. This is the mass compatible with relativistic mean field theory of nuclear matter. The other is $\sim 700$ MeV which figures as an effective scalar meson in chiral Lagrangian with the parameters scaling with density~\cite{song}. Given the uncertainty, these values should be taken as simply representative. For definiteness, we will focus more on $m_\chi\approx 700$ MeV.

The results of \cite{PKR} essential for what follows are:
\begin{enumerate}
\item
The state of skyrmions in FCC makes a phase transition to a half-skyrmion matter in CC at $n=n_{1/2} > n_0$ (where $n_0\approx 0.16$ fm$^{-3}$ is the normal nuclear matter density).  It is found to be fairly independent of the mass of the scalar $\chi$~\cite{LPRV}: Even at 1200 MeV, it differs negligibly from that of 700 MeV. However it is sensitive to the Skyrme parameter $e$ going as $\sim e^3$ -- and more generally to certain HLS parameters such as the HLS coupling $g$, so it is difficult to pin down the density $n_{1/2}$. For the given $f_\pi$ and $e$, it comes at the range $(1.3-2)n_0$. This should be taken as representative. What is important for our purpose is that the $n_{1/2}$ be not too far above $n_0$. Were it to be so, then the effect of the phase change would be unimportant in the process we are concerned with.

We note that in this phase, the quark condensate $\la\bar{q}q\ra^*\propto ({\rm Tr} U)^*$ vanishes {\it on averaging} while the average value of the amplitude field remains non-zero.
\item
As density is increased beyond $n_{1/2}$, a phase change takes place at $n_c$ to a matter where $f_\pi^*$ drops to zero. The phase change appears to be of first order. The critical density $n_c$ for this phase with $\la\bar{q}q\ra^*=f_\pi^*=0$ turns out to be extremely sensitive to the dilaton mass. For instance, $n_c/n_0 \approx (4-25)$ for $m_\chi\approx (700-1200)$ MeV~\cite{LPRV}. This phase could be identified as the chiral-symmetry restored and quark-deconfined phase.
\end{enumerate}

\section{New Scaling}
From the above results, we infer the following consequences on in-medium scaling.

\subsection{Scaling of the nucleon mass}

Within the range of density involved, the large $N_c$ piece of the effective (or quasi-)nucleon mass in the model scales as $m_N^*\sim f_\pi^*/e$. Since $e$ is scale-invariant, the scaling is only in $f_\pi^*$. As noted, $f_\pi^*$ remains non-zero, with the ratio $f_\pi^*/f_\pi$ dropping roughly linearly in density to a non-zero value at $n_c$~\cite{footnote2}. To a good first approximation, we can simply take
\be
m_N^*/m_N&\approx& f_\pi^*/f_\pi \ \ \ {\rm for}\ \ 0\lsim n \lsim n_{1/2} \label{Nscaling1}\\
&\approx& b \ \ \  \ \ \  \ \ \ {\rm for} \ \ n_{1/2}\lsim n \lsim n_c
\ee
where $b$ is a constant, a reasonable range of which is $b\sim 0.6-0.8$. What we have here resembles the result obtained in the parity-doubled linear or nonlinear sigma model where the chiral invariant mass $m_0$ comes out to be $m_0\sim 0.5 - 0.8$ MeV~\cite{parity-doubling}.
\subsection{Scaling of the vector-meson mass}

For the properties of vector mesons, we shall be guided by the HLS theory to which (\ref{skyrme-lag}) is gauge-equivalent~\cite{hls,HY:PR}. We expect the vector-meson mass to take the form (in the leading $N_c$ order)
\be
m_V^*\approx f_\pi^* g_V^*
\ee
where $g_V^*$ is the hidden gauge coupling constant and $V=\rho, \omega$. We have set $a^*=(f_\sigma^*/f_\pi^*)^2=1$ for large $N_c$. There is no theoretical argument as to how $g_V^*$ scales in density up to $n_0$ (or $n_{1/2}$). However thermal lattice calculations indicate that there is practically no scaling up to near the critical temperature. In \cite{BR:DD}, this observation was simply carried over to the density case. We were led to assume that $g_V^*$ does not scale up to $n_{1/2}$. In \cite{BR:DD}, this argument is given some support from nuclear dynamics. However the RG argument based on hidden local symmetry theory shows that as one approaches the chiral critical point $n_c$, it should scale to zero proportionally to $\la\bar{q}q\ra$ very near the VM fixed point~\cite{HY:PR}. Thus we infer the following scaling:
\be
\frac{m_V^*}{m_V}&\approx&  f_\pi^*/f_\pi\equiv \Phi \ \ \ {\rm for} \ \  0\lsim n \lsim n_{1/2}\label{Phi}\\
&\approx& b \frac{g_V^*}{g_V}\equiv b\Phi^\prime \ \  {\rm for} \ \  n_{1/2}\lsim n \lsim n_c\label{Phip}
\ee
where the two different scaling functions $\Phi$ and $\Phi^\prime$ are defined.
It is important to note that the vector-meson mass scales with $f_\pi^*$ up to $\sim n_{1/2}$ but with $g_V^*$ for $n>n_{1/2}$. This observation was made already in \cite{BR:DD} and applied to kaon condensation in \cite{kaoncond}. It is the scaling (\ref{Phi}) that is revealed in the C14 dating~\cite{holt}. In hidden local symmetry theory without baryon degrees of freedom~\cite{HY:PR}, $g_V^*$ scales $\propto \la\bar{q}q\ra^*$ but given the presence of the half-skyrmion phase, this scaling could be modified. What is certain is that it will go to zero at the VM fixed point even in the presence of baryons.

\section{Effect on Nuclear Tensor Forces}
We shall now apply the scaling relations (\ref{Nscaling1})-(\ref{Phip}) to nuclear tensor forces. That the scaling proposed in \cite{BR91} could strongly affect the nuclear tensor forces  in nuclear matter has been known since some time~\cite{BR:DD}. The new scalings leave the behavior up to $n_0$ unchanged from the previous scaling but drastically modify the properties above the nuclear matter density.

Since the nucleon is massive in our model in the whole range of density we are dealing with, the non-relativistic approximation is valid, so the two-body tensor forces contributed by one pion and one $\rho$ exchange maintain the familiar form
 \begin{eqnarray}
V_M^T(r)&&= S_M\frac{f_{NM}^2}{4\pi}m_M \tau_1 \cdot \tau_2 S_{12}\nonumber\\
&& \left(
 \left[ \frac{1}{(m_M r)^3} + \frac{1}{(m_M r)^2}
+ \frac{1}{3 m_Mr} \right] e^{-m_M r}\right),
\label{tenforce}
\end{eqnarray}
where $M=\pi, \rho$, $S_{\rho(\pi)}=+1(-1)$. The key aspect of these forces is that there is a strong cancelation between the two.
This cancelation plays the crucial role in the C12 dating problem in \cite{holt}.

Thus far, we have not addressed how pion properties scale. If the pion mass were zero, they would be protected by chiral invariance and hence would remain unscaled. But with non-zero pion mass, the situation could be different. Since the chiral symmetry is only lightly broken, the in-medium property of the pion is subtle and requires an extremely careful treatment. Such an analysis by Jido, Hatsuda and Kunihiro yielded the in-medium  Gell-Mann-Oakes-Renner relation~\cite{jidoetal}
\be
m^*_\pi(n)/m_\pi \approx (f_\pi^t (n)/f_\pi)^{-1}(\la\bar{q}q\ra^*(n)/\la\bar{q}q\ra)^{1/2}
\ee
where $f_\pi^t$ is the time component of the pion decay constant which differs from the space component in medium. Using the experimental information available at the nuclear matter density~\cite{hatsuda-hayano}, $(f_\pi^t (n_0)/f_\pi)^2\simeq 0.64$ and $\la\bar{q}q\ra^* (n_0)/\la\bar{q}q\ra\simeq 0.63$, we get $m_\pi^*/m_\pi\simeq 1$. So there is no noticeable change in the pion mass up to the nuclear matter density. It seems reasonable to assume that up to the density we are concerned with -- which is not too far above $n_0$ --, the pion properties remain unscaled.  This should be good enough for our discussion. Taking into account the small pion mass effect in a more precise way would require a highly detailed treatment that the model used here is not equipped to handle -- and that is not warranted for the qualitative aspect we are exploring in this paper.

To see how the $\rho$ tensor force scales, we need to see how the strength $f_{N\rho}$ scales. Plugging in the vector (or hidden gauge) coupling $g_V$, the strength scales as
\be
R\equiv \frac{f_{N\rho}^*}{f_{N\rho}}\approx \frac{g_V^*}{g_V}\frac{m_\rho^*}{m_\rho}\frac{m_N}{m_N^*}.
\ee
It follows from the scaling relations (\ref{Nscaling1})-(\ref{Phip}) that
\be
R&\approx& 1\ \ \ {\rm for}\ \ \ 0\lsim n \lsim n_{1/2}\label{R1} \\
& \approx& {\Phi^\prime}^2 \ \ {\rm for} \ \  n_{1/2}\lsim n \lsim n_c.\label{R2}
\ee

In \cite{holt}, Eqs. (\ref{Phi}) and (\ref{R1}) were used to explain the long life time in the C14 dating beta decay. The change of scaling that takes place at $n > n_{1/2}$ was not probed in that process.
Were one to extend the scalings (\ref{Phi}) and (\ref{R1}) employed by \cite{holt} to higher densities, one would find that the net tensor force would be nearly completely suppressed for the inter-nucleon separation $r\gsim 1.5$ fm~\cite{xu-bal} at $n\sim 3n_0$. However, this behavior is drastically  modified for $n\gsim n_{1/2}$ by the change of the scaling (\ref{R2}). Because of the strong quenching of the $\rho$ tensor strength while the mass drops, the cancelation between the two tensor forces gets abruptly weakened as density passes $n_{1/2}$. In fact, with $\Phi^\prime$ (assumed to be) scaling linearly in density, the $\rho$ tensor force gets more or less completely killed at $n\gsim 2n_0$, leaving only the $\pi$ tensor operative, in a stark contrast to the naive scaling that suppresses the total tensor instead.

\begin{figure}[ht!]
\begin{center}
\includegraphics[height=7.cm]{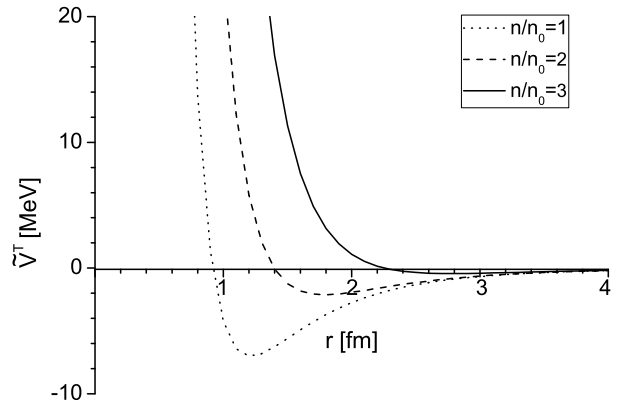}
\vskip -0.5cm
\caption{Sum of $\pi$ and $\rho$ tensor forces $\tilde{V}^T\equiv (\tau_1 \cdot \tau_2 S_{12})^{-1} (V_\pi^T +V_\rho^T)$ in units of MeV for densities $n/n_0$ =1, 2 and 3 with the ``old scaling," $\Phi \approx  1-0.15 n/n_0$ and $R\approx 1$ for all $n$.}\label{tensorold}
\end{center}
\end{figure}

\begin{figure}[ht!]
\begin{center}
\includegraphics[height=7.cm]{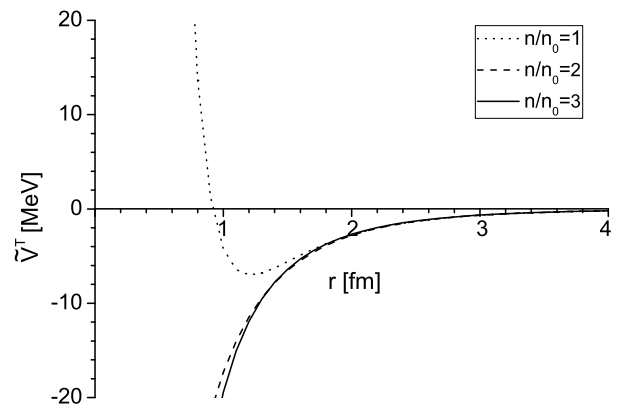}
\vskip -0.5cm
\caption{The same as Fig.~\ref{tensorold} with the ``new scaling," $\Phi \approx 1-0.15 n/n_0$  with  $R\approx 1$ for $n<n_{1/2}$ and $R\approx \Phi^2$ for $n>n_{1/2}$, assuming $ n_0<n_{1/2}<2n_0$.}\label{tensornew}
\end{center}
\end{figure}

To give a quantitative idea of the drastic change that takes place due to the half-skyrmion phase, we compare the behavior of the tensor forces as a function of density between the ``old" and ``new" scalings. They are given in Figs.~\ref{tensorold} and ~\ref{tensornew} for the choices of the scalings indicated therein. For simplicity of illustration, we took $b\approx 1$ and $\Phi=\Phi^\prime\approx 1-0.15n/n_0$. Given that the scalings $\Phi$ and $\Phi^\prime$ are unknown for $n>n_0$, what we have gotten here is qualitative at best. Numerically the results are not sensitive to the value of $b$ near 1, but they could depend quantitatively on the way that $\Phi$ and $\Phi^\prime$ scale.

\section{Symmetry Energy}
The strong suppression of the $\rho$ tensor will clearly have a big effect on the structure of baryonic matter at high density. We test this feature in the nuclear symmetry energy that figures importantly in the structure of neutron-rich nuclei and more crucially in neutron stars. The most dramatic effect can be illustrated with the ``supersoft" symmetry energy recently discussed in \cite{xu-bal,bal-fopi,bal-nonnewton}.

\begin{figure}[hbt]
\includegraphics[width=0.4\textwidth,angle=-90]{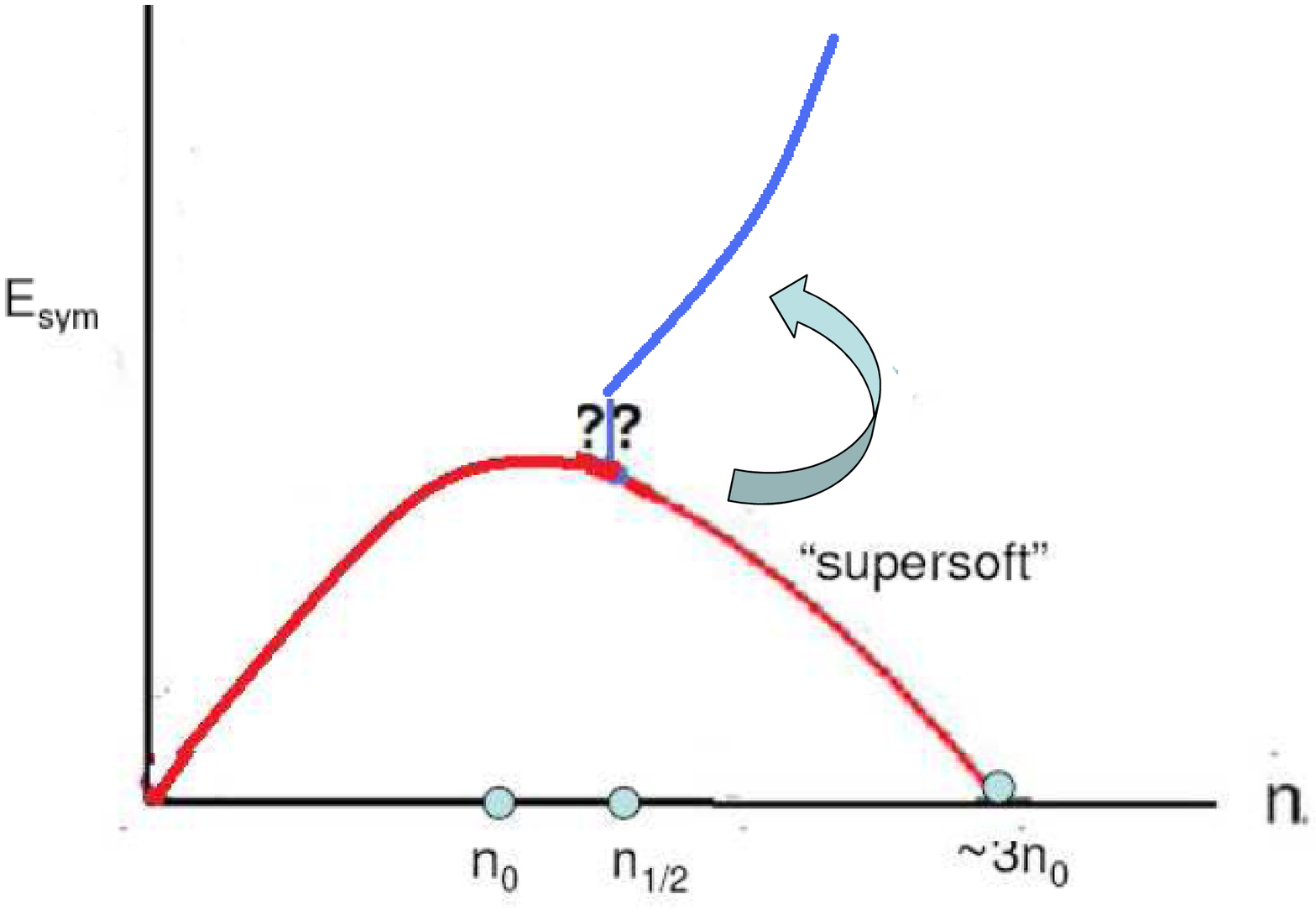}
\vskip -0.5cm
\caption{A cartoon of the symmetry energy. The scaling operative in the half-skyrmion phase brings the abrupt change from attraction to repulsion as indicated by the arrow. The change-over region indicated by ?? is not understood as described in Fig.~2 of \cite{BR:DD}.}
\label{schematic}
\end{figure}

The energy per particle of asymmetric nuclear matter is given by
\be
E(n,\delta)=E_0(n)+E_{sym}(n) \delta^2 +\cdots
\ee
 where $\delta=(N-P)/(N+P)$ with $N(P)$ the number of neutrons (protons) and the ellipsis stands for higher orders in $\delta$. We focus on the ``symmetry energy" $E_{sym}$. Fit to experimental data up to $n_0$, most of the symmetry energy predicted theoretically are found to increase monotonically up to $n_0$ with, however, a wide variation above $n_0$ due to the paucity of experimental constraints and the lack of reliable theory. In \cite{bal-fopi}, what we might refer to as a ``non-standard form" of $E_{sym}$ that increases up to, and turns over at, $\sim n_0$, deviating from the standard form and vanishing near $3n_0$ is argued to be required by the FOPI/GSI data on $\pi^-/\pi^+$ data. The schematic form of such $E_{sym}$ is shown as ``supersoft" in Fig.~\ref{schematic}. It is immediately clear that such a non-standard symmetry energy will have dramatic consequences in nuclear physics, astrophysics and other areas. For instance, it would modify the Newtonian gravity~\cite{bal-nonnewton}, falsify the scenario of kaon condensation in compact-star formation and collapse to black holes~\cite{BB} etc. Whether or not such a supersoft $E_{sym}$ (SSE for short) is picked by Nature is going to be tested in forthcoming experiments at a variety of laboratories such as RIB, FAIR/GSI etc. Leaving that issue to the future, let us take the SSE as an illustration of an extreme case and ask whether and how our scaling enters into the structure of that symmetry energy. In fact we were motivated to ask this question by the work of Xu and Li~\cite{xu-bal} who have shown that if the tensor forces with the scaling (\ref{Phi}) and (\ref{R1}) {\it applied to all densities} (or alternatively three-body forces) were taken into account, then the ``standard symmetry energy" that increases in density continuously could be made to turn over at $\sim n_0$ and take the form of the SSE.

The key mechanism exploited in \cite{xu-bal} is the cancelation taking place {\em at and beyond} $n\gsim n_0$ in the tensor forces. Our prediction, on the contrary, is that the cancelation in question will cease precisely when the skyrmion fractionizes into half-skyrmions at $n_{1/2}$. Given the abrupt suppression of the $\rho$ tensor, we expect that the curve will turn over from decreasing to increasing at $n_{1/2}$. Our expectation based on the new scaling is schematically shown in a cartoon in Fig.~\ref{schematic}. As stressed in \cite{BR:DD}, it is not known how the changeover takes place. It may or may not have a discontinuity, but our analysis suggests that the slopes before and after $n_{1/2}$ will differ.

\section{Symmetry Energy in Half-Skyrmion Matter}
In order to make Fig.~\ref{schematic} realistic so as to apply to neutron-star systems, one would have to
formulate a microscopic approach to implement the tensor forces with the predicted scaling in both $E_0$ and $E_{sym}$. It may require implementing three-body forces as well. It will be a highly involved calculation that would require a lot more work; it is being pursued at present~\cite{wcu-texas}.

Here we would like to suggest that the new scaling can be ``seen" directly from the half-skyrmion matter on which our scaling relations are based. In the skyrmion framework, the symmetry energy comes from a term subleading in $N_c$. It must therefore arise from the collective quantization of multi-skyrmion systems. In \cite{kopelio}, the Skyrme model (supplemented with a six-derivative term but without the dilaton field) was collective-quantized to obtain the Weizs\"acker-Bethe-Bacher formula for neutron-rich (even and odd A) nuclei from $A=6$ to $A=32$. The symmetry energy so obtained is in good agreement with experimental spectra. This suggests using the same technique to compute the symmetry energy from the skyrmion crystal. In his original work on the skyrmion crystal, Klebanov~\cite{klebanov} discussed how to collective-quantize the pure neutron system. We apply this method to the skyrmion matter as well as to the half-skyrmion matter we have obtained.

Consider an $A$-nucleon system for $A\rightarrow\infty$. Following Klebanov, the whole matter is rotated through a single set of collective
coordinates $U(\vec{r}, t) = A(t) U_0(\vec{r}) A^\dagger (t)$ where $U_0(\vec{r})$ is the static crystal configuration with the
lowest energy for a given density. The canonical quantization leads to
\be
E^{\mbox{tot}} = A M_{\mbox{cl}}
+ \frac{1}{2A \lambda_{I}} I^{\mbox{tot}} (I^{\mbox{tot}}+1),
\ee
where $M_{\mbox{cl}}$ and $\lambda_{I}$ are, respectively, the mass and the moment of inertia
of the single cell.
The moment of inertia is of the form
\begin{eqnarray}
&&\lambda_I = \int_{\mbox{Cell}} d^3 x \left\{
\frac{f_\pi^2}{6} \left(\frac{\chi}{f_\chi}\right)^2
\textstyle (3-\frac{1}{2}\mbox{tr} (U_0 \tau_a U_0^\dagger \tau_a )) \right.
\nonumber\\
&& \displaystyle +\frac{1}{24e^2} \left[
\textstyle (3-\frac{1}{2}\mbox{tr} (U_0 \tau_a U_0^\dagger \tau_a ))
\mbox{tr} (\partial_i U_0^\dagger \partial_i U_0 ) \right.
+\mbox{tr} (\partial_i U_0 \tau_a \partial_i U_0^\dagger \tau_a )
\nonumber\\
&& \left.\left. \textstyle
+ \frac12 \mbox{tr} (\partial_i U_0 U_0^\dagger \tau_a \partial_i U_0 U_0^\dagger \tau_a)
+ \frac12 \mbox{tr} (\partial_i U_0^\dagger U_0 \tau_a \partial_i U_0^\dagger U_0 \tau_a)
\right]\right\}.\nonumber
\end{eqnarray}
$I^{\mbox{tot}}$ is the total isospin which would be the same as the third component of the isospin $I_3$ for pure neutron matter. This suggests taking for $\delta\equiv (N-P)/(N+P)\lsim 1$
\be
I^{\mbox{tot}}=\frac 12 A\delta.
\ee
Thus the energy per nucleon in an infinite matter ($A=\infty$) is
\be
E=E_0 +\frac{1}{8\lambda_I}\delta^2.\label{E}
\ee
with $E_0=M_{cl}$.
This leads to the symmetry energy
\be
E_{sym}=\frac{1}{8\lambda_I}.
\ee

\begin{figure}[hbt]
\includegraphics[width=0.35\textwidth,angle=-90]{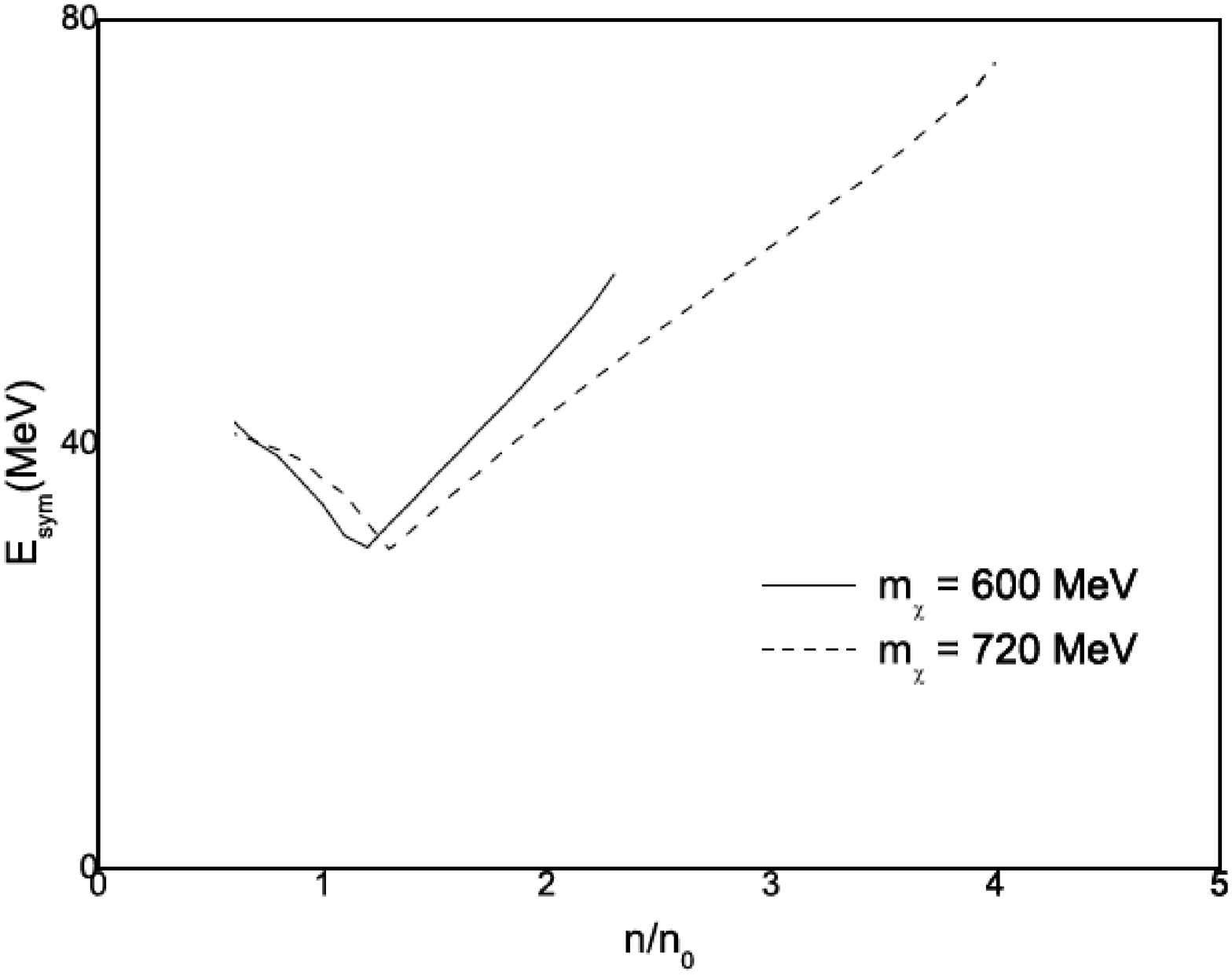}
\vskip -0.cm
\caption{Symmetry energy given by the collective rotation of the skyrmion matter with $f_\pi=93$ MeV, $1/e^2\approx 0.03$ and two values of dilaton mass. The cusp is located at $n_{1/2}$. The low density part that cannot be located precisely is not shown as the collective quantization method used is not applicable in that region.}
\label{skyrme-SE}
\end{figure}

The numerical results for the given parameters are plotted for densities below and above $n_0$ in Fig.~\ref{skyrme-SE}. The Klebanov collective quantization method is not expected to be applicable for very low densities. The refined treatment made in \cite{kopelio} with the Skyrme model (using the rational-map approximation) of mass splittings of nuclear isotopes shows that the $E_{sym}$ in finite systems decreases as the mass number $A$ increases. This implies that the decrease in $E_{sym}$ seen just below $n_0$ in Fig.~\ref{skyrme-SE} is consistent with the result of \cite{kopelio}. The striking feature, i.e., the cusp at $n_{1/2}$, reproduces what is expected with the new scaling relations (\ref{Phip}) and (\ref{R2}) as schematically shown in Fig.~\ref{schematic}.

One might raise an objection at this point on the reliability of calculating the symmetry energy $E_{sym}$ as a collective quantization correction. With the Lagrangian (\ref{skyrme-lag}), it is very likely that the symmetric part of the energy of the nucleonic matter, i.e., $E_0$ in (\ref{E}), will not saturate at the correct density. This is because there is not enough repulsion in the model that would balance the attraction coming from the dilaton, i.e., the mechanism that reduces the soliton mass to $\sim 1$ GeV in \cite{PRV-dilaton}, near the saturation density.

We would argue that this problem does not affect our conclusion. Clearly the $\omega$ field, when implemented as in \cite{kopelio}, will remove this defect. What we are calculating for $E_{sym}$ is a $1/N_c$ effect like the $N-\Delta$ mass difference since it is $\propto 1/\lambda_I$ and the moment of inertia $\lambda_I\propto N_c$. What enters into $\lambda_I$ is the leading $N_c$ term and the subleading effects that figure in the saturation in $E_0$ would not affect $E_{sym}$ to the leading order we are considering. An evidence for this is the fact that the location of the density $n_{1/2}$ is extremely insensitive to the dilaton mass as one can see in Fig.~3 of \cite{LPRV} whereas the $E_0$ -- hence the density $n_\chi$ at which $\la\bar{q}\ra^*=f_\pi^*=0$ -- is strongly affected by the dilaton mass.

\section{Further comments}

We should mention that the model used in this article has several caveats that need to be addressed. For instance, large $N_c$ arguments invoked for the crystal structure of dense matter may be invalidated by $1/N_c$ (quantum) corrections. The quantum fluctuations could melt the crystal, turning the skyrmion crystal into a skrymion liquid. Zahed discussed what happens to the dyonic salt crystal in holographic QCD when the system is heated, and arrived at a low temperature $\sim 10$ MeV for the system to melt into a dyonic liquid~\cite{zahed}. Quantum fluctuation and thermal fluctuation are expected to act in a similar way. The question would be whether this melting invalidates the argument for the change of the scaling in the density regime we have considered and its impact on the tensor forces. We have no clear answer to this.  However we would conjecture that since the transition from the skyrmion matter to the half-skyrmion matter is topological, the qualitative feature would survive owing to topological stability.

A plausible consequence of the new scaling proposed in this paper is the formation of a neutron solid with $\pi^0$ condensate discussed a long time ago~\cite{pi0}. In \cite{pi0}, Pandharipande and Smith argued that with certain enhancement of the pion tensor force, the crystal structure with $\pi^0$ condensate should be more favored energetically than liquid structure. Since our new scaling makes the $\rho$ tensor suppressed while the pion tensor is left strong at high density, the half-skyrmion phase in solid form could be {\em a posteriori} justified.

It seems possible that a continuum description of the half-skyrmion matter suitable for liquid structure at moderate density -- prior to the possible $\pi^0$ condensation -- is related to Georgi's vector symmetry conjectured to be present in the large $N_c$ limit~\cite{georgi}. That the chiral symmetry is putatively restored with ${\rm Tr}U=0$ but the pion decay constant remains non-zero may be interpreted in hidden local symmetry theory in terms of $f_\pi\neq 0$ and $a=1$. We should point out that it also resembles the ``hadronic freedom" regime invoked for the region of hot/dense matter between the ``flash point" -- at which hadrons in medium go $\sim$ 90 \% on-shell -- and the chiral transition point in which the vector coupling $g_V\approx 0$, with the hadronic interactions becoming weak~\cite{brownetal-dilepton}. This issue calls for a rigorous treatment.

Also, the Skyrme model supplemented with the soft dilaton only and without the $\omega$ meson degree of freedom -- which figures as a six-derivative term in \cite{kopelio} -- may be too simplistic. Some of those degrees of freedom that are integrated out -- such as the tower of vector mesons including the $\omega$ meson -- may have to be considered explicitly. The extreme sensitivity of the phase transition point $n_c$ -- in contrast to $n_{1/2}$ -- to the dilaton mass may be a signal for this. Nonetheless, the picture we have obtained for the tensor forces and the symmetry energy seems to be consistent and qualitatively robust. This may have to do with the topological nature of the transition involved as in certain condensed matter processes.

The present formulation offers a possibility of determining the symmetry energy for neutron-star matter even with hyperons present. This could be done by collective-quantizing multi kaons bound in the skyrmion matter constrained with beta equilibrium. This would also allow one to study dense multi-kaonic nuclear matter along the line discussed in \cite{PKR} for one anti-kaon.

We stress that the forthcoming experiments at RIB, FAIR etc., could check the anomalous behavior of the symmetry energy at $n_{1/2}$ and if present, determine $n_{1/2}$. It is amusing to note that such measurements would also pin down the constant $e$ and possibly other parameters if introduced in the effective chiral Lagrangian. If it turned out that $n_{1/2}$ were not too high above $n_0$, it should be of no exaggeration to state that the scaling proposed here would have strong consequences on {\em all} processes that probe highly compressed cold matter, say, at FAIR.
\vskip 0.5cm

\subsection*{Acknowledments}
We are grateful for useful correspondence with Bao-An Li and Chang Xu on the symmetry energy, specially on their work in \cite{xu-bal}.
This work was supported by the WCU project of Korean
Ministry of Education, Science and Technology (R33-2008-000-10087-0).


\begin{thebibliography}{99}
\bibitem{PKR} B.~Y.~Park, J.~I.~Kim and M.~Rho,
  ``Kaons in dense half-skyrmion matter,''
  Phys.\ Rev.\  C {\bf 81}, 035203 (2010)
  [arXiv:0912.3213 [hep-ph]].
\bibitem{RSZ}  M.~Rho, S.~J.~Sin and I.~Zahed,
  ``Dense QCD: a holographic dyonic salt,''
  Phys.\ Lett.\  B {\bf 689}, 23 (2010)
  [arXiv:0910.3774 [hep-th]].

\bi{SS} T.~Sakai and S.~Sugimoto,
``Low energy hadron physics in holographic QCD,''
 { Prog.\ Theor.\ Phys.}\  {\bf 113}, 843 (2005);
    [arXiv:hep-th/0412141];
    ``More on a holographic dual of QCD,''
 { Prog.\ Theor.\ Phys.}\  {\bf 114}, 1083 (2005).
  [arXiv:hep-th/0507073].

\bi{ice9} It was first pointed out to one of us (MR) by Gerry Brown that this phenonomenon is somewhat like the ``Ice-9" in K. Vonnegut, {\it Cat's Craddle}\ (Holt, Rinehardt \& Winston, USA, 1963).
\bi{LR-dilaton} H.~K.~Lee and M.~Rho,
  ``Dilatons in hidden local symmetry for hadrons in dense matter,''
  Nucl.\ Phys.\  A {\bf 829}, 76 (2009).
  [arXiv:0902.3361 [hep-ph]].

\bi{hls} M. Bando, T., Kugo, and K. Yamawaki,
``Nonlinear realization and hidden local symmetries,"
{Phys. Rept.} {\bf 164}, 217 (1988).


\bi{SS-son} D.T. Son and M.A. Stephanov,
``QCD and dimensional deconstruction,"
Phys. Rev. {\bf D69}, 065020 (2004).

\bi{diakonov} For details, see D. Diakonov and V. Petrov,
``Nucleons as chiral solitons,"
in {\it  At the
frontier of particle physics: Handbook of QCD}\ ed by M. Shifman
(World Scientific, Singapore, 2001) \, vol. 1, pp. 359-415.

\bi{footnote0} The hQCD model~\cite{SS} gives $1/e^2\approx 2.51 \frac{\lambda N_c}{216\pi^3}\approx 0.02$ for $N_c=3$ and $\lambda\approx 17$ that is fit by the meson and baryon porperties. This is comparable to what one obtains using the mass formula $m_\rho^2=2f_\pi^2 g_V^2\approx 770$ MeV, i.e., $1/g_V^2\approx 0.03$.

\bi{footnote-integrating} It has recently been shown~\cite{HMY} that if all vector mesons {\em except} for the lowest -- $V_0= (\rho, \omega)$ -- are integrated out in a way consistent with hidden local symmetry for {\em all} vector mesons in the Sakai-Sugimoto model~\cite{SS}, the resulting Lagrangian for $V_0$ (with the Goldstone $\pi$) is precisely the HLS theory proposed in \cite{hls,HY:PR}, with the parameters of the Lagrangian fixed by the 5D holographic QCD. This Lagrangian is found to give a new interpretation of vector dominance for both the pion~\cite{HMY} and the nucleon~\cite{HR} in their electromagnetic form factors. It is also interesting to note that the Skyrme Lagrangian with only the current algebra term and the Skyrme quartic term is obtained (in the bulk sector) when {\em all} vector mesons are integrated out in the hidden local symmetric way. This can be interpreted as the statement (in the gauge sector) of ``gauge equivalence"  between nonlinear sigma model and HLS theory~\cite{hls,HY:PR}.

\bi{HMY} M.~Harada, S.~Matsuzaki and K.~Yamawaki,
  ``Integrating out holographic QCD back to hidden local symmetry,''
  arXiv:1003.1070 [hep-ph].

\bi{HY:PR} M.~Harada and K.~Yamawaki,
  ``Hidden local symmetry at loop: A new perspective of composite gauge boson
  and chiral phase transition,''
  Phys.\ Rept.\  {\bf 381}, 1 (2003).

 \bi{HR} M.~Harada and M.~Rho,
  ``Holographic vector dominance for the nucleon,''
  arXiv:1010.1971 [hep-ph].





\bi{NRZ} M.~A.~Nowak, M.~Rho and I.~Zahed,
  {\it Chiral Nuclear Dynamics}\
{(World Scientific, Singapore, 1996)}

\bi{footnote1} The $N_c$ counting goes as follows: $\sim$ 1500 MeV at ${\cal O}(N_c)$, $\sim -500$ MeV at ${\cal O}(N_c^0)$ and $\sim 300$ MeV at ${\cal O}(1/N_c)$ (from $N-\Delta$ mass difference). We see nothing unreasonable in this counting.

\bibitem{BR91}
  G.~E.~Brown and M.~Rho,
  ``Scaling effective Lagrangians in a dense medium,''
  Phys.\ Rev.\ Lett.\  {\bf 66}, 2720 (1991).


\bi{PRV-dilaton} B.~Y.~Park, M.~Rho and V.~Vento,
  ``The role of the dilaton in dense skyrmion matter,''
  Nucl.\ Phys.\  A {\bf 807}, 28 (2008).
  [arXiv:0801.1374 [hep-ph]].

\bi{footnote1p} In the hQCD model~\cite{SS}, it is the $U(1)$ field in the Chern-Simons term -- and not the Skyrme quartic term -- that stabilizes the instanton. It is possible that this role is played by the $\omega$ field also in the HLS model.

\bi{footnoteattempt} This problem was addressed in \cite{PRV-dilaton} with an HLS Lagrangian in unitary gauge. However the analysis made there used an approximation which is most likely invalid in dense medium: there the anomalous Lagrangian consisting of four homogeneous Wess-Zumino (hWZ) terms was approximated by only one term proportional to $\omega_\mu B^\mu$ where $B^\mu$ is the baryon current. The assumption there was that the vector meson stays ``heavy" at any density, which is at odds with the VM property. A correct calculation with all four hWZ terms remains to be done.

\bi{klebanov} I.~R.~Klebanov,   ``Nuclear matter in the Skyrme model,''
  Nucl.\ Phys.\  B {\bf 262}, 133 (1985).

\bi{byp-vv} {\it The Multifaceted Skyrmion}\  ed. by G.E. Brown and M. Rho (World Scientific Publishing, Singapore, 2010).

\bi{song}  C.~Song,
  ``Dense nuclear matter: Landau Fermi-liquid theory and chiral Lagrangian
  with scaling,''
  Phys.\ Rept.\  {\bf 347}, 289 (2001).

\bi{LPRV} H.~J.~Lee, B.~Y.~Park, M.~Rho and V.~Vento,
  ``The pion velocity in dense skyrmion matter,''
  Nucl.\ Phys.\  A {\bf 741}, 161 (2004)
  [arXiv:hep-ph/0307111].


\bi{footnote2} This observation is consistent with the pion decay constant ``measured" in deeply bound pionic nuclei, $f_\pi^*/f_\pi\approx 0.8$ at the nuclear matter density. In finite nuclei there are nuclear corrections subleading in $N_c$ that we are not considering. Note that in HLS theory~\cite{HY:PR}, it is the loop corrections higher order in $1/N_c$ that drive the physical (renormalized) pion decay constant to zero at the chiral transition.

\bi{parity-doubling} D.~Zschiesche {\it et al.},
  ``Cold, dense nuclear matter in a SU(2) parity doublet model,''
  Phys.\ Rev.\  C {\bf 75}, 055202 (2007);
  S.~Gallas, F.~Giacosa and D.~H.~Rischke,
  ``Vacuum phenomenology of the chiral partner of the nucleon in a linear sigma
  arXiv:0907.5084 [hep-ph].

\bi{BR:DD}  G.~E.~Brown and M.~Rho,
  ``Double decimation and sliding vacua in the nuclear many-body system,''
  Phys.\ Rept.\  {\bf 396}, 1 (2004)
  [arXiv:nucl-th/0305089].

\bi{kaoncond} G.~E.~Brown, C.~H.~Lee, H.~J.~Park and M.~Rho,
  ``Study of strangeness condensation by expanding about the fixed point of
  the Harada-Yamawaki vector manifestation,''
  Phys.\ Rev.\ Lett.\  {\bf 96}, 062303 (2006)
  [arXiv:hep-ph/0510073].


\bi{holt} J.~W.~Holt {\it et al.}, \
  ``Shell model description of the $^{14}$C dating beta decay with Brown-Rho-scaled
  NN interactions,''
  Phys.\ Rev.\ Lett.\  {\bf 100}, 062501 (2008)
  [arXiv:0710.0310 [nucl-th]].

\bi{jidoetal} D.~Jido, T.~Hatsuda and T.~Kunihiro,
  ``In-medium pion and partial restoration of chiral symmetry,''
  Phys.\ Lett.\  B {\bf 670}, 109 (2008)
  [arXiv:0805.4453 [nucl-th]].
\bi{hatsuda-hayano} R.~S.~Hayano and T.~Hatsuda,
  ``Hadron properties in the nuclear medium,''
  arXiv:0812.1702 [nucl-ex].


\bi{xu-bal} C. Xu and B.-A. Li, arXiv:0910.4803.
\bi{bal-fopi} Z.G. Xiao {\it et al.}, Phys. Rev. Lett. {\bf 102}, 062502 (2009).
\bi{bal-nonnewton} D.-H. Wen, B.-A. Li and L.-W. Chen, Phys. Rev. Lett. {\bf 103}, 211102 (2009).

\bi{BB} G.E. Brown and H.A. Bethe,
``A scenario for a large number of low-mass black holes in
the galaxy,"
{ Astrophys. J.} {\bf 423},  659 (1994).

\bi{wcu-texas} K. Kim, H.K. Lee and M. Rho, work in progress
\bi{kopelio} V.B. Kopeliovich, A.M. Shunderuk and G.K. Matushko, Phys. of Atom. Nuclei {\bf 69}, 120 (2006).

\bi{LPMRV}  H.~J.~Lee, B.~Y.~Park, D.~P.~Min, M.~Rho and V.~Vento,
  ``A unified approach to high density: Pion fluctuations in skyrmion
  matter,''   Nucl.\ Phys.\  A {\bf 723}, 427 (2003).

\bi{pi0} V.R. Pandahripande and R.A. Smith, ``A model neutron solid with $\pi^0$ condensate," Nucl. Phys. {\bf A237}, 507 (1975).

\bi{zahed} I.~Zahed,
  ``Holographic nucleons,''
  arXiv:1010.5980 [hep-ph].

\bi{georgi} H. Georgi,
``New realization of chiral symmetry,"
{Phys.\ Rev.\ Lett.}\ {\bf 63}, 1917 (1989).

\bi{brownetal-dilepton} G.E. Brown {\it et al.}, ``Hidden local field theory and dileptons in relativistic heavy ion
  collisions,''
  Prog.\ Theor.\ Phys.\  {\bf 121}, 1209 (2009)
  [arXiv:0901.1513 [hep-ph]].


\end{thebibliography}
\end{document}